\documentclass[11pt,a4paper]{article}
\usepackage{jcappub}

\title{Rotation in the Geometric Scalar Theory of Gravity}

\author[a,b]{Mario Novello}
\author[a,b]{and Vicente Antunes,}
 \affiliation[a]{Centro de Estudos Avan\c{c}ados de Cosmologia (CEAC-CBPF), Rua Dr. Xavier Sigaud 150, Urca, CEP 22290-180,
 Rio de Janeiro, RJ, Brazil}
\affiliation[b]{Centro Brasileiro de Pesquisas F\'{i}sicas (CBPF), Rua Dr. Xavier Sigaud 150, Urca, CEP 22290-180,\\
 Rio de Janeiro, RJ, Brazil}

\emailAdd{novello@cbpf.br}
\emailAdd{antunes@cbpf.br}

\arxivnumber{ }

\abstract{We present solutions corresponding to rotational configurations in the recently proposed Geometric Scalar Gravity (GSG) theory. The solutions obtained here have the important property that the associated closed time-like curves 
 are always restricted to a compact domain of space-time. We compare this property with those of analogous solutions in General Relativity (GR).}

\begin{document}

\maketitle

\section{Introduction}

In the first attempts to formulate a relativistic theory of gravitation made by Nordstr\"{o}m \cite{nordstrom}, Einstein and Grossmann \cite{einsteingrossmann} in the early 1910's, the gravitational field was described by the special relativistic generalization of the Newtonian potential, which was assumed to propagate in a conformally flat space-time and to obey a generalised Poisson equation with the trace of the energy-momentum tensor of matter as its source.
  The failure of that program on both theoretical and observational grounds opened the way for the emergence of General Relativity (GR) and its rapid progression towards its current status as the paradigmatic theory of gravity. During the last decades, however, GR itself has been facing a growing pressure in both observational and theoretical fronts, a situation that has prompted the search of alternatives to GR. Recently, a theory of gravitation based on a single scalar field that avoids the drawbacks of those early formulations was proposed by Novello \textit{et al.} \cite{novelloetal}. In particular, unlike Nordstr\"{o}m and Einstein-Grossmann proposals, the gravitational interaction in this new approach --- called Geometric Scalar Gravity (GSG) --- is described in terms of a modified space-time geometry, in accordance with the fundamental principle of GR. Furthermore, the source of the scalar field in GSG is not simply the trace of the energy momentum-tensor of matter. Some cosmological and astrophysical scenarios within the GSG framework were studied in a series of papers \cite{bittencourtetal, moschellaetal}.

The purpose of the present work is to present novel geometries in the framework of GSG corresponding to rotational configurations. It is well known that many solutions to Einstein field equations displaying rotational effects, such as Lanczos-van Stockum dust, G\"{o}del's rotating universe and Taub-NUT space-time, exhibit closed timelike curves (CTC's) that are not hidden by any horizon, a feature that implies (global) causality violations. Furthermore, most of these solutions cannot be ruled out on theoretical grounds, a fact that exposes the incompleteness of GR and its need to be complemented by chronology protection conjectures. 

 Here, we analyze two rotational configurations in GSG: a vacuum solution corresponding to a compact domain, called I, and that corresponding to an exterior domain filled with a perfect fluid, named domain II.
It is shown that in the interior (vacuum) solution, there exists CTCs which are constrained to stay within the compact domain I ($ r < r_{m}.$). Besides, we show that there are no CTCs in the outside metric, \textit{i.e.} in domain II. We compare this property with a similar situation that occurs in Gödel's solution.

\section{A brief review of Geometric Scalar Gravity \label{GSG}}

The GSG theory is based on the following postulates:

\begin{itemize}
	\item{The gravitational interaction is described by a
		scalar field $ \Phi$;}
	\item{The field $\Phi$ satisfies a nonlinear dynamics;}
	\item{The theory satisfies the principle of general covariance which,
		in other words, means that GSG is not a theory restricted to the realm of special relativity;}
	\item{All kinds of matter and energy interact with $\Phi$
		only through the pseudo-Riemannian metric
		\begin{equation}
			q^{\mu\nu} = \alpha \eta^{\mu\nu} + \frac{\beta}{w} 
			\partial^{\mu}\Phi \partial^{\nu} \Phi,
			\label{4jul}
		\end{equation}
		where $ \alpha$ and $\beta$ are functions of $\Phi$ and $w \equiv
		\partial_{\mu}\Phi \partial_{\nu} \Phi \eta^{\mu\nu};$}
	\item{Test particles follow geodesics relative to the gravitational metric $ q_{\mu\nu};$}
	\item{ $\Phi$ is related in a nontrivial way with the Newtonian potential $\Phi_N$;}
	\item{The background Minkowski metric $ \eta^{\mu\nu}$ is not observable. Matter and energy interact gravitationally only
		through the combination 
		\[ 
		\alpha \eta^{\mu\nu} + \frac{\beta}{w} \partial^{\mu}\Phi \partial^{\nu} \Phi
		\] 
		and its derivatives;}
	\item{Electromagnetic waves propagate along null geodesics relative to the metric $ q^{\mu\nu};$}
	\item{The contravariant definition of the metric in GSG, that is,
		equation (\ref{4jul}) is an exact expression. The corresponding
		covariant expression, defined as the inverse $q_{\mu\nu} q^{\nu\lambda} = \delta^{\lambda}_{\mu}$, is also a binomial expression
		\begin{equation}
			q_{\mu\nu} = \frac{1}{\alpha}  \eta_{\mu\nu} - \frac{\beta}{\alpha (\alpha + \beta) w} \partial_{\mu} \Phi \partial_{\nu}
			\Phi, \label{291} \end{equation} }
\end{itemize}
where $ w = \partial_{\mu}\Phi \partial_{\nu}\Phi \eta^{\mu\nu}.$
\vspace{0.5cm}
 
 	According to these postulates, the action for the matter fields can be written as
 	\begin{equation}
 	S_{m} =\int \sqrt{-q}\, L_m\, d^4x,
 	\end{equation}
 	where  $q$ is the determinant of $q_{\mu\nu}$ and we are taking $\kappa = c= 1$. The dynamics of the scalar field (see \cite{novelloetal}) is given by 
   \begin{equation}\label{eqgsg}
   	\sqrt{{V}}  \square\Phi= \kappa \chi,
   	\end{equation}
   where
   \begin{equation}
   	\chi\equiv \,-\,\frac{1}{2}\left[\,T +\left(2 -\frac{V'}{2V}\right)E +C^\lambda_{~\,;\lambda}\right],
   \end{equation}
   \begin{equation}
   	E \equiv \frac{T^{\mu\nu} \partial_{\mu}\Phi \partial_{\nu}\Phi}{\Omega},
   \end{equation}
   \begin{equation}
   	C^{\lambda}\equiv\frac{(\alpha^2V-1)}{\Omega} \, \left( T^{\lambda\mu} - E q^{\lambda\mu} \right) \partial_{\mu}\Phi.
   \end{equation}
   Here, $T$ is the trace of the energy-momentum tensor, $V' = dV/d\Phi$ and $\Omega = \partial_{\mu}\Phi \partial_{\nu}\Phi q^{\mu\nu} = (\alpha + \beta) w.$
   In the case examined in \cite{novelloetal} $ \alpha = \exp (- 2 \Phi) $  and the values of $\beta$ and $ V$ are given by
   	\begin{equation*}
 	4 \, \alpha^3 \, V = (\alpha - 3)^2,
 	\end{equation*}
   \begin{equation*}
 	\alpha + \beta = \alpha^3 V.
 	\end{equation*}

   Note that the vacuum dynamics of $ \Phi$ can be written equivalently in the unobservable Minkowski background by the action
   \begin{equation}
   	S_{gravity} = \int \sqrt{- \eta} \,\frac{V(\Phi)}{w}  \,\partial_{\mu}\Phi \, \partial_{\nu}\Phi \,  \, \eta^{\mu \nu}.
   \end{equation}

\section{Rotation in Vacuum} \label{rotation}

Let us write the line element associated with the background metric in the form

\begin{equation}	
ds^2_{M} = \eta_{\mu\nu} dx^{\mu} dx^{\nu} = dt^2 + 2 h(r) \, dt d \phi - dr^2 - dz^2 + g(r) \, d\phi^2. \label{metric}
\end{equation}	
In the case where the functions $g(r)$ and $h(r)$ are given by
$$ h(r) = h_{0} , \hspace{0.50cm} g(r) = h_{0}^2 - w_{0}^2 \, r^2,$$
where $h_{0}$ and $w_{0}$ are constants, the metric (\ref{metric}) is nothing but the flat Minkowski geometry. The $q$-metric from equation (\ref{4jul}) leads to the associated line element for the foreground geometry
\begin{align}
ds^2 & = q_{\mu\nu}dx^{\mu}dx^{\nu} \nonumber\\
& = \frac{1}{\alpha} \, dt^2 - \frac{1}{(\alpha + \beta)} \, dr^2 + \frac{(h_0^2 - w^2 r^2)}{\alpha} \,d\phi^2 - \frac{1}{\alpha} \, dz^2 \nonumber\\
& + \frac{2 h_0}{\alpha} \, dt \, d\phi. \label{qmetric}
\end{align}

 \subsection{Vacuum solution}

Let us consider the vacuum dynamics of the scalar field 
\begin{equation}
	\Box \,  \Phi
	= \frac{1}{\sqrt{- q}}
	\partial_{\mu} ( \sqrt{- q} \,q^{\mu\nu} \partial_{\nu} \Phi
	)=0,  \label{kkgg}
\end{equation}
In the case in which $ \Phi$ depends only on the variable $r$, the field that satisfies equation (\ref{qmetric}) is given by
\begin{equation}
	\frac{(\alpha - 1)}{\alpha^{3/2}} = b \,\log r
\end{equation}
where $b$ is a constant. For simplicity we take $ b =1.$ Note that $r$ as an inverse function of the field $\alpha$ has a maximum value for the point $\alpha =3$ which is the singular boundary of this geometry.

Note however that this geometry is restrained to the values of r: $0 < r < r_{m}$ where the maximum radius is given by $r_{m} = \exp{(2 \sqrt{3}/9)}.$ This value of $r,$  corresponds to the value $ \alpha = 3,$ where the metric is singular. Indeed, the scalar of curvature is given by
$$ R = \frac{14 \, \alpha^3  + 18 \alpha}{(\alpha - 3)^2 w_{0}^2 \, r^2}. $$

\subsection{CTC\lowercase{s} \label{ctcs}}

Let us consider a curve $ x^\mu (s)$ such that $z = constant, t = constant, r = constant,$ 
that is on the curve we have
\begin{equation}
 ds^2 = \frac{1}{\alpha} \, \left( h_{0}^2 - w^2 \, r^2 \right) \, d\phi^2 .
\end{equation}
 Defining $r_c^2 = h_0^2/w_{0}^2$, we can see that this closed curve has the following property:
 \begin{itemize}
 	\item{For $r > r_{c} $ the curve is space-like;}
 	\item{For $r < r_{c}$ the curve is time-like.}
  \end{itemize}

We note that this situation is distinct from that encoded by others metrics, like for instance that of Gödel\rq s geometry, which allows the presence of CTCs in an exterior region of a domain defined by a given radius $r_G$. Here, on the contrary, the CTCs are confined in the interior domain of radius $r_{m}.$ This property precludes the application of Gauss' prescription, by means of which it is always possible to construct a Gaussian system of coordinates separating the 3-d space from time (at least, in a small domain).

 \section{Dynamics in the presence of matter \label{matter}}

From what we have shown in the previous section, it is natural to try to avoid the singularity of this metric by joining this geometry to another one in the exterior domain, before the value $r_m$ is reached. In order to do this, consider a fluid  with density $ \rho$ without pressure, so that the energy-momentum tensor admits the expression
 \begin{equation*}
 T^{\mu\nu} = \rho \, v^{\mu} \, v^{\nu}.
 \end{equation*}
 We set $ v^{\mu} = ( A, 0, B, 0)$, and normalize this vector by imposing the relation between $A$ and $B$
 \begin{equation}
 \alpha = A^2 + h_{0} \, A B + ( h_{0}^2 - w_{0}^2 \, r^2) \, B^2.
 \end{equation}
 This vector will be a geodesic under the condition
 \begin{equation}
 \frac{d\alpha}{dr} + 2 w_{0}^2 \, r \, B^2 = 0,
 \end{equation}
 where we have used the normalization condition for the vector $v^{\mu} $.
 
 The Geometric Scalar Gravity theory provides the dynamics of the metric as
 \begin{equation}
 	\sqrt{V} \, \Box \Phi = \kappa \, \chi ,
 	\label{1481}
 \end{equation} 
 where in this case 
\begin{equation*}
\chi = \frac{1}{2} \, \rho.
\end{equation*}
 It then follows 
 \begin{equation}
 	\frac{1}{16} \, ( \alpha - 3)^2 \, \frac{1}{r} \, \frac{d}{dr} \Delta= -\frac{1}{2} \, \rho,
 \end{equation}
 where
\begin{equation}
\Delta \equiv   (\alpha - 3) \, \alpha^{-5/2} \,r \,  d\alpha/dr.
\end{equation}

  Let us consider the case in which $ \Delta = 2\lambda/r^2.$ Thus the density of energy is given by 
  \begin{equation}
  	\frac{\lambda}{4} \, \frac{(\alpha - 3)^2}{r^3} = - \, \frac{\kappa}{2} \, \rho.
  	\end{equation}
  The expression of $\Delta$ yields
  \begin{equation}
  	\frac{(\alpha - 1)}{\alpha^{3/2}} = \frac{\lambda}{r}.
  \end{equation}
 
 Note that this geometry is assymptotically  flat. Indeed, when $ r $ goes to infinity the density of energy vanishes and the function $\alpha$ goes to $ 1,$ that is, the q-metric becomes the Minkowski metric. Note that once the constant $\lambda$ is negative, the value of $\alpha$ in the domain II must be bounded: $ 0 < \alpha < 1.$ 
 
 Te next step is to join the metrics of domains I and II in order to avoid the presence of singularity at $ \alpha - 3 = 0.$ It then follows for the value $r_{b} < r_{m}$ the condition
 \begin{equation}
 \exp {(\lambda/r_{b})} = r_{b}.
 	\end{equation}

\section{Final comments \label{final}}

 The property displayed by the metric of the domain I shows a very unusual feature: the existence of CTCs in a compact domain. This is not the case in the most famous solution of GR found by Kurt Gödel \cite{godel}, \cite{novellogodel} . Indeed, Gödel\rq s geometry written under the form of the metric (8) has the functions $h$ and $g$ given by  
 $$ h = \sqrt{2} \, \sinh^2 r ,$$
 $$g = 2 \sinh^4 r - \sinh^2 r \cosh^2 r .$$
  Define the critical radius $ r_{c}$ of Gödel\rq s metric by setting 
 $\sinh r_{c} = 1.$ For values of $r$ such that $ 0 < r < r_{c} $ the value of $g$ is negative. However, for values of $ r $ such that $ r > r_{c}$ the curve is time-like, that is a CTC. Thus, for Gödel\rq s metric there are no CTCs in the interior domain $ r <  r_{c}$. CTCs appear only in the exterior domain of G\"{o}del's metric, $ r> r_{c}$. As we have shown above, this behaviour is opposite to the one obtained for a rotational configuration in GSG theory.

 \section{Acknowledgements}

  The authors would like to thank the Brazilian agencies Faperj (MN), CNPq, and the CBPF (VA) for their financial support.

\end{document}